# Dichromatic "breather molecules" in a mode-locked fiber laser

Yudong Cui[1,3,5*], Yusheng Zhang[2], Lin Huang[4], Aiguo Zhang[4], Zhiming Liu[4], Cuifang Kuang[1,3], Chenning Tao[2], Daru Chen[2], Xu Liu[1,3], and Boris A. Malomed[6,7]

*1 State Key Laboratory of Modern Optical Instrumentation, College of Optical Science and Engineering, Zhejiang University, Hangzhou 310027, China*

*2 Hangzhou Institute of Advanced Studies, Zhejiang Normal University, Hangzhou, China*

*3 ZJU-Hangzhou Global Scientific and Technological Innovation Center, No.733 Jianshe San Road, Xiaoshan District, Hangzhou, Zhejiang, China, 311200*

*4 Ceyear Technologies Co., Ltd, Qingdao 266555, China*

*5 Wuhan National Laboratory for Optoelectronics, Huazhong University of Science and Technology, Wuhan, 430074 China*

*6 Department of Physical Electronics, Faculty of Engineering, and Center for Light-Matter Interaction, Tel Aviv University, Tel Aviv 69978, Israel*

*7 Instituto de Alta Investigación, Universidad de Tarapacá, Casilla 7D, Arica, Chile*

*These authors contributed equally: Yudong Cui, Yusheng Zhang*

*\*Corresponding author: cuiyd@zju.edu.cn*

**Abstract：** Bound states of solitons ("molecules") occur in various settings, playing an important role in the operation of fiber lasers, optical emulations, encoding, and communications. Soliton interactions are generally related to breathing dynamics in nonlinear dissipative systems, maintaining potential applications in spectroscopy. In the present work, dichromatic breather molecules (DBMs) are created in a synchronized mode-locked fiber laser. Real-time delay-shifting interference spectra are measured to display the temporal evolution of the DBMs, that cannot be observed by means of the usual real-time spectroscopy. As a result, robust out-of-phase vibrations are found as a typical intrinsic mode of DBMs. The same bound states are produced numerically in the framework of a model combining equations for the population inversion in the mode-locked laser and XPM-coupled complex Ginzburg-Landau equations for amplitudes of the optical fields in the fiber segments of the laser cavity. The results demonstrate that the Q-switching instability induces the onset of breathing oscillations. The findings offer new possibilities for the design of various regimes of the operation of ultrafast lasers.

The generation of solitons plays a profound role in optics, fluids, plasmas, Bose-Einstein condensates, and other fields [1–7]. This concept was further extended to nonlinear dissipative media [3,7–9], where solitons are supported by balance of gain and loss [10–14]. A versatile platform to realize various species of dissipative solitons is provided by mode-locked fiber lasers [14–18]. In the lasers [13,14,19], as well as in microresonators and passive fiber cavities [20–22], dissipative solitons generally exhibit breathing behavior. Soliton creeping [23,24] and subharmonic entrainment were also reported for the breathers [25,26]. This phenomenology is related to the Fermi-Pasta-Ulam recurrence, rogue waves, and modulation instability [27–29]. In addition to their significance to fundamental studies, breather solitons offer applications to dual-comb spectroscopy and design of supercontinuum sources [30,31].

Interactions between solitons feature various outcomes, including quasi-elastic collisions and formation of "soliton molecules" (SMs) [4,29,32-43], so called due to similarities of their properties to those of atomic molecules, such as synthesis and intrinsic vibrations [33,44]. SMs can carry a quaternary code, thus offering applications to optical communications [45]. Vibration dynamics of SMs has also been studied in detail [14,29,42,46]. Benefitting from the development of real-time spectroscopy, transient interactions, periodic vibrations, and formation of multi-soliton structures have been observed, restoring temporal-domain patterns on the basis of spectral-interference ones [4,13,47–50].

Copropagating modes with different carrier wavelengths form polychromatic SMs [51–53]. Models of dichromatic SMs with quartic group-velocity dispersion (GVD) [52,54–56] and in micro-resonators [56,57] were recently elaborated. However, intrinsic vibrations of polychromatic SMs have not yet been observed. This is a challenge for real-time spectroscopy, as bound states of solitons with different wavelengths do not produce interference patterns. In the present work, dichromatic breather molecules (DBMs) are created and quantified, and temporal oscillations in them are studied by dint of real-time delay-shifting interference spectra (RDIS). The experiment is performed in a passive synchronously mode-locked fiber laser, with the pulse trapping maintained via cross-phase modulation (XPM) [58,59]. RDIS makes it possible to record interference patterns of adjacent pulses, determined by phase and position shifts between them. Thus, trajectories of the breathers paired at different wavelengths, as well as the separation and phase difference between them, can be obtained. Further, numerical simulations are implemented to reproduce the experimentally observed DBM dynamics.

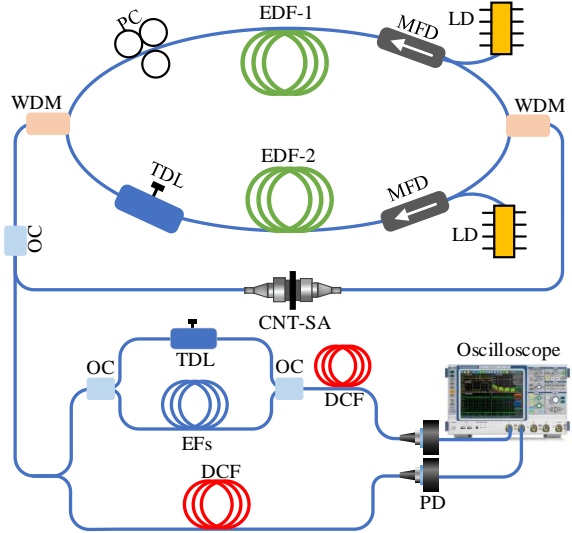

Figure 1. The experimental setup built for studying DBMs.

The experimental setup is shown in Fig. 1. Two distinct paths for optical signals carried by wavelengths 1550 and 1570 nm are produced by two wavelength-division multiplexers (WDMs). Each path includes an erbium-doped fiber (EDF), a polarization controller (PC), and a multifunctional device (MFD) that combines the pump with the signal and maintains the unidirectional operation. A time-delay line (TDL) is utilized to adjust the path difference for compensating the GVD-induced delay. The saturable absorber (SA) and output coupler (OC) are shared by both paths. The GVD is compensated by TDL, making the optical paths for the dichroic solitons nearly identical. The solitons are thus created under the condition of the quasi-group-velocity matching, cf. Refs. [51–57,60]. Real-time spectra are recorded by temporally stretching the solitons in a 5-km-long GVD-compensating fiber (DCF). The output pulses enter an interferometer, where an extra fiber is added in one branch to build a bound state of two adjacent pulses. The net GVD is adjusted to make it nearly identical in both branches.

The so constructed SM is stretched by the dispersion fiber to monitor real-time spectra. This technique was used to study subharmonic entrainment of breather solitons [26]. Here, RDIS is used to extract the temporal structure of DBM as follows. Denoting the shift of the temporal position of the breather produced by an $R$-th roundtrip (RT) in the circuit as $f(R)$, the temporal separation in the SM can be written as $\Delta\tau(R) = f(R) - f(R-1) + T_0$, $R = 2, 3, 4\ldots, N$, where $T_0$ is the pulse separation in the SM built by the interferometer. It can be calculated as the average separation for DBMs, assuming that their intrinsic vibrations are quasi-harmonic. Eliminating $T_0$, the separation is given by $\Delta\tau_g(R) = f(R) - f(R-1)$. If the initial temporal position is set as $f(1)=0$, the temporal position of the breather after $N$ RTs is

$$f(N) = \sum_{R=2}^{N} \Delta\tau_g(R), N \geq 3. \tag{1}$$

As a result, the evolution of the relative temporal position can be obtained from the interference fringes of the SM. Then, the relative motion of the DBMs is identified as $\Delta f(R) = f_1(R) - f_2(R)$, where $f_1(R)$ and $f_2(R)$ are temporal positions of the two breathers. The phase shift between them is identified similarly.

In the experiment, static dichromatic SMs have been generated with a suitable pump power, time delay and polarization state, as shown in Figs. 2(a) and 2(b). They exhibit two humps in the spectral domain and the interference pattern in temporal one, forming a meta-envelope, which is a characteristic feature of dichromatic SMs [51-56]. By slightly decreasing the pump power, DBM is achieved. The spectral evolution measured by the real-time spectroscopy is displayed in Fig. 2(c), where periodic variations of the dichroic components are observed with respect to the number of RTs. The real-time spectra reveal identical frequency-time relationship for the two solitons, implying that they are mutually trapped by XPM. The corresponding evolution of the autocorrelation trace exhibits the interference pattern for each wavelength, which is a hallmark of the dichromatic SM. The period of the observed temporal modulation is the inverse of the frequency difference corresponding to the two wavelengths.

The 3-dB bandwidth ($\Delta\lambda_{3\mathrm{dB}}$) and central wavelength ($\lambda_\mathrm{c}$) of the breathers are obtained from Fig. 2(c). Here, the centroid method is employed to identify $\lambda_\mathrm{c}$ of the asymmetric spectrum. The central wavelengths vary periodically as shown in Fig. 2(d), with the corresponding frequency shift induced by the XPM, third-order GVD, and self-phase modulation of the pulses, cf. Refs. [61–63]. In Fig. 2(e), bandwidths $\Delta\lambda_{3\mathrm{dB}}$ for the two carrier wavelengths also display periodic evolution, and, accordingly, pulse widths demonstrate the breathing behavior produced by the anomalous GVD. The breathing ratios, of the maximum bandwidth to the minimum one, are 1.10 and 1.13 for the two frequencies, respectively. Thus, the breathing modes are weak ones, with asymmetric spectra. Note that the evolution of $\Delta\lambda_{3\mathrm{dB}}$ and $\lambda_\mathrm{c}$ for the two breathers feature the same period, 15 RTs, but are mutually out-of-phase, in contrast with the usual dynamics of XPM-trapped pulses [64].

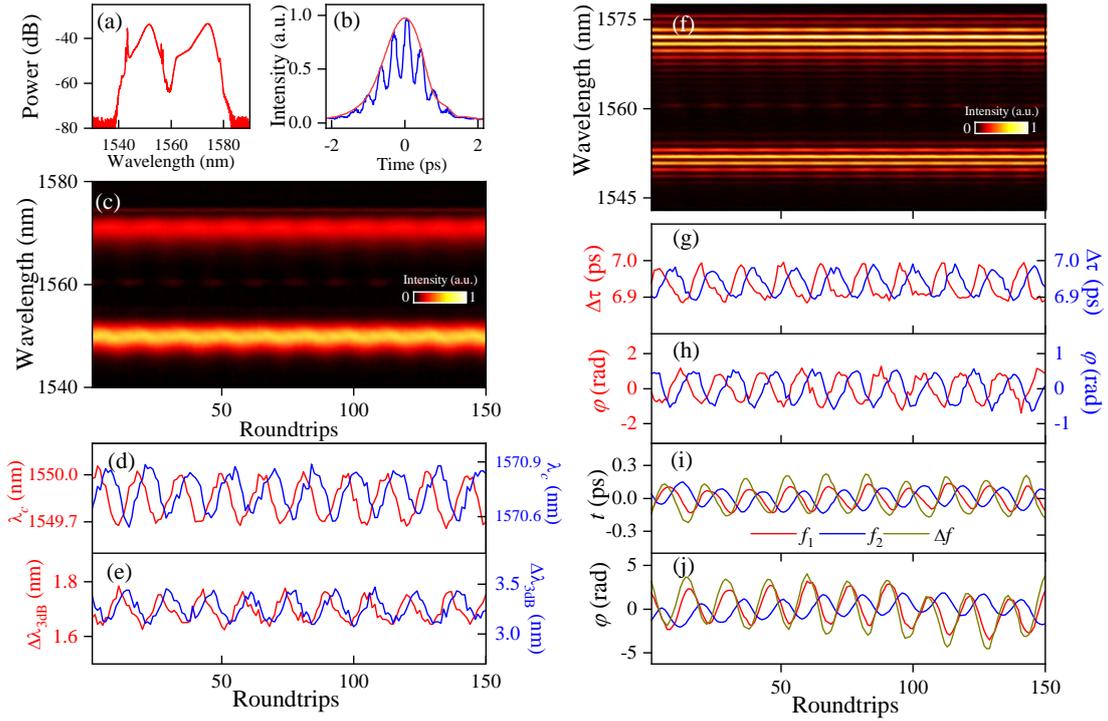

Fig. 2. Experimental results for DBMs. (a) The optical spectrum of the static dichromatic SM. (b) The autocorrelation trace, with the red line showing the meta-envelope. (c) The real-time spectral evolution. (d) The evolution of the central wavelength and (e) the 3-dB bandwidth calculated from (a). (f) The interference spectral pattern produced by RDIS. The SM with different wavelengths is built,

exhibiting periodic variations. (g) The evolution of the pulse separation and (h) phase difference, as extracted from (f). (i) The calculated vibrational dynamics of DBM and trajectories of the two breathers. (j) The evolution of the phase difference of DBM and phases of the two breathers. The red and blue curves pertain to the breathers carried by wavelengths 1550 nm and 1570 nm, respectively.

Figure 2(c) does not reveal the relative motion and phase variation between the breathers which form the DBM. This dynamics, as produced by RDIS, is presented in Fig. 2(f), where spectral fluctuations imply variation of the separation and phase shift of SMs. The pulse separation ($\Delta\tau$) and phase shift ($\varphi$) of the SMs can be extracted from the spectral interference patterns shown in Figs. 2(g) and 2(h), that exhibit the same modulation period as Figs. 2(d) and 2(e). Then, the trajectory of motion ($f(R)$) and phase evolution ($\Phi(R)$) of the two breathers can be calculated by accumulating $\Delta\tau$ and $\varphi$, as shown in Fig. 2(i) and 2(j). Taking the difference of $f(R)$ and $\Phi(R)$ for the breathers, the relative motion and phase evolution of the DBM constituents can be inferred. The separation between the paired breathers exhibits oscillations with amplitude $\approx 0.343$ ps. The dynamical variables, including $\Delta\lambda_{3\mathrm{dB}}$ and $\lambda_c$, as well as $f(R)$ and $\Phi(R)$ for the same wavelength, oscillate out-of-phase, which, as said above, is a characteristic feature of DBMs.

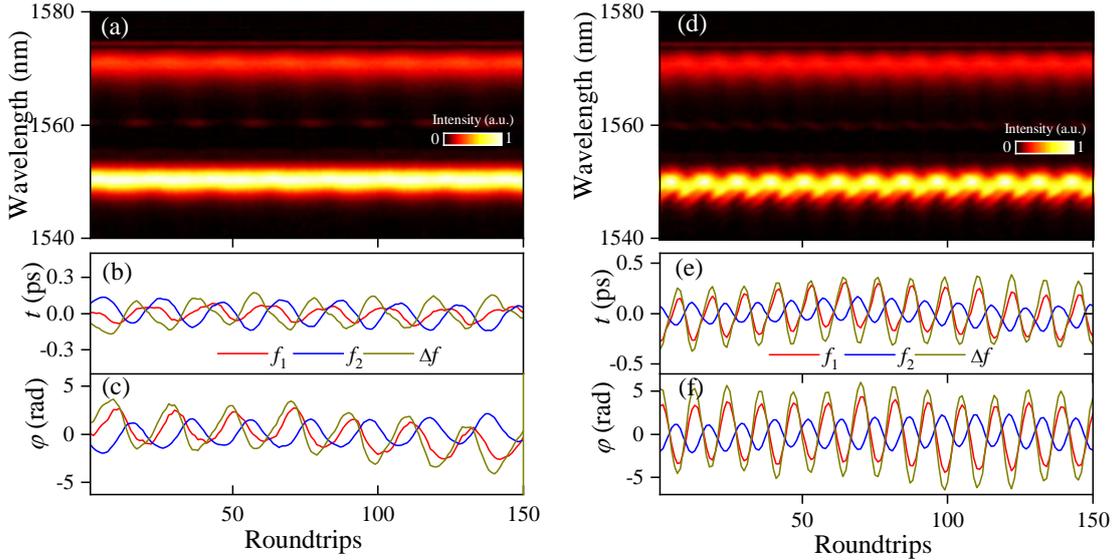

Fig. 3. DBMs with different vibration amplitudes. A nearly static SM: (a) the real-time spectral evolution; (b) time-domain dynamics; (c) the phase evolution. Panels (d,e,f):

the same for SM with enhanced vibrations.

For different pump powers and optimized polarization states, two other characteristic cases are shown in Fig. 3, to present the results in a broader form. When the pump power supplied by one LD is slightly increased, the resultant picture is displayed in Fig. 3(a), and details produced by RDIS are presented in Fig. S3 of Supplementary Material [65]. The corresponding breathing ratio is ~1.08, which is smaller than that in Fig. 2. Accordingly, the DBM vibrations are weaker, with amplitude $\approx 0.269$ ps, and the phase variation is smaller too, as shown in Figs. 3(b) and 3(c). The breathing behavior is enhanced if both pump powers are reduced, leading to DBMs with breathing ratios $\approx 1.253$ and $1.138$, and the vibration amplitude $\approx 0.646$ ps, as shown in Figs. 3(d) and 3(e). Thus, the DBM vibrational dynamics is controlled by the pump powers, which has a simple explanation: the soliton's effective mass is proportional to its energy [1], i.e., to the pump power. Then, the amplitude of mutual oscillations of the bound solitons is inversely proportional to their mass, according to the mechanical laws of motion.

The experimental results for the DBMs, summarized above, were compared to numerical simulations. The propagation of DBMs in the fiber segments and shared single-mode fiber between two WDMs (Fig. 1) is governed by CGLEs (2) and (3), respectively, for slowly varying amplitudes $u_1$ and $u_2$ of the pulses at the carrier wavelengths 1550 nm and 1570 nm [7,8,66,67]:

$$\frac{\partial u_{1,2}}{\partial z} = \frac{g}{2} u_{1,2} - i \frac{\beta_2}{2} \frac{\partial^2 u_{1,2}}{\partial t^2} + i\gamma |u_{1,2}|^2 u_{1,2} + \frac{g}{2\Omega_g^2} \frac{\partial^2 u_{1,2}}{\partial t^2}, \qquad (2)$$

$$\frac{\partial u_{1,2}}{\partial z} = -\frac{1}{v_{1,2}} \frac{\partial u_{1,2}}{\partial t} - i \frac{\beta_2}{2} \frac{\partial^2 u_{1,2}}{\partial t^2} + i\gamma \left( |u_{1,2}|^2 + 2|u_{2,1}|^2 \right) u_{1,2}. \qquad (3)$$

Here $t$ and $z$ are the time and propagation distance, $v_{1,2}$ are group velocities at the two wavelengths, $\beta_2$ and $\gamma$ represent the GVD and nonlinearity of the fiber, while $g$ and $\Omega_g$ are the gain strength and spectral bandwidth. However, the time delay based on group velocities also originates from the dispersion in fiber. So the group-velocity mismatch should be eliminated, if the time delay is applied to two solitons via $\beta_2$ in Eqs. (2). The distribution of the signal and pump intensities, $I_s(z)$ and $I_p(z)$, along EDF is governed by the two-level rate equations [66,67]:

$$\frac{dI_{s,p}(z)}{dz} = \Gamma_{s,p} \left[ \sigma_{s,p}^{(e)} N_2 - \sigma_{s,p}^{(a)} N_1 \right] I_{s,p}(z),$$ (4)

$$\frac{dN_{1,2}}{dt} = \pm\Gamma_{21}N_2 \pm \left[ \sigma_s^{(e)} N_2 - \sigma_s^{(a)} N_1 \right] \frac{I_s}{\hbar\omega_s} \pm \left[ \sigma_p^{(e)} N_2 - \sigma_p^{(a)} N_1 \right] \frac{I_p}{\hbar\omega_p},$$ (5)

Here $\sigma_{p,s}^{(a/e)}$ are the absorption/emission cross sections for the pump and signal at 980 nm and 1560 nm, respectively, $N_{1,2}$ represent population densities of the ground and excited states, $\Gamma_{21}=1/\tau$ is the probability of the spontaneous transition between them, and $\Gamma_{s,p}$ are modal overlap factors. $N_{1,2}$ can be calculated, as functions of $z$ in the course of the RT time, from Eq. (5), the total population being $N = N_1 + N_2$. The intensities of the signal and pump at given $z$ can be obtained from Eq. (4), and these values are used to calculate the population at $z + dz$. Solution of Eqs. (4) and (5) produce the gain coefficient as $g(z) = (I_s(z))^{-1}dI_s(z)/dz$ [12,13]. The parameters of the gain medium is as following: $\sigma_s^{(e)}$=5.3×10$^{-25}$ m$^2$, $\sigma_s^{(a)}$=3.5×10$^{-25}$ m$^2$, $\sigma_p^{(a)}$=3.2×10$^{-25}$ m$^2$, $N$=5.4×10$^{24}$ m$^{-3}$, $\tau$=12 ms, $\Gamma_{s,p}$=0.4.

Next, the action of the SA (see Fig. 1) is provided by a transfer function (TF), $|u_{1,2}|^2 \rightarrow [1-\alpha_0/(1+P/P_{sat})] |u_{1,2}|^2$, where $\alpha_0$ is the modulation depth, $P$ the instantaneous power, and $P_{sat}$ the saturation power. The TDL is represented by the corresponding phase delay, $u_1 \rightarrow u_1\exp(-i\omega t_D)$, where $\omega$ is the optical frequency, and $t_D$ the time delay between the two branches. Finally, the 1550/1570 WDM is provided by two bandpass filters with TFs $F_{1,2} = \exp\left(-(1/2)\left[(\omega-\omega_{1,2})/\Delta\omega_{1,2}\right]^8\right)\exp\left(-i(\omega-\omega_{1,2})^3\beta_3\right)$, where $\omega_{1,2}$ are the respective carrier frequencies, $\Delta\omega_{1,2}$ are bandwidths of the filters, and $\beta_3$ is the third-order GVD introduced by them. The full solution obtained for an RT provides the input for the next RT. The simulations were initiated with a weak Gaussian pulse (the peak power less than $1\times10^{-9}$ W). We used the measured values for dispersion $\beta_2$, and the calculated nonlinear coefficients $\gamma = 4.5$ W$^{-1}$km$^{-1}$ and 1.3 W$^{-1}$km$^{-1}$ for EDF and SMF, respectively. We used the following parameters to produce the DBM: $t_D = -0.26321$ ps, $\beta_3 =$ -0.005 ps$^3$ for 1550 nm and $\beta_3 =$ -0.035 ps$^3$ for 1570 nm; $\alpha_0 = 0.2$, $P_{sat} = 30$ W; $\Omega_g = 40$ nm, $\Delta\omega_{1,2}$=10 nm.

In comparison to previous works, where the population is determined adiabatically by the signal and pump powers, the present system demonstrates modulations of the pulse intensity induced by the Q-switching-driven instability, cf. Refs. [68,69]. It originates from the SA operation at the pump power below the onset of the saturation, which is identical to the mechanism for the generation of breathers in mode-locked fiber lasers [13]. Irrespective of a weak input, the output converges to a stable or periodically evolving state that is independent of the initial signal. An example of the numerically generated DBM is presented in Fig. 4. In Fig. 4(b) an oscillatory bound state of two breathers, overlapping in the temporal domain, produces interference fringes at each carrying wavelength, as seen in Fig. 4(a), which shows temporal profiles of both breathers. The corresponding spectral evolution, as predicted by the theoretical model, is plotted in Fig. 4(c). Similar to its experimental counterpart in Fig. 2(a), it demonstrates breathing dynamics at both carrying wavelengths.

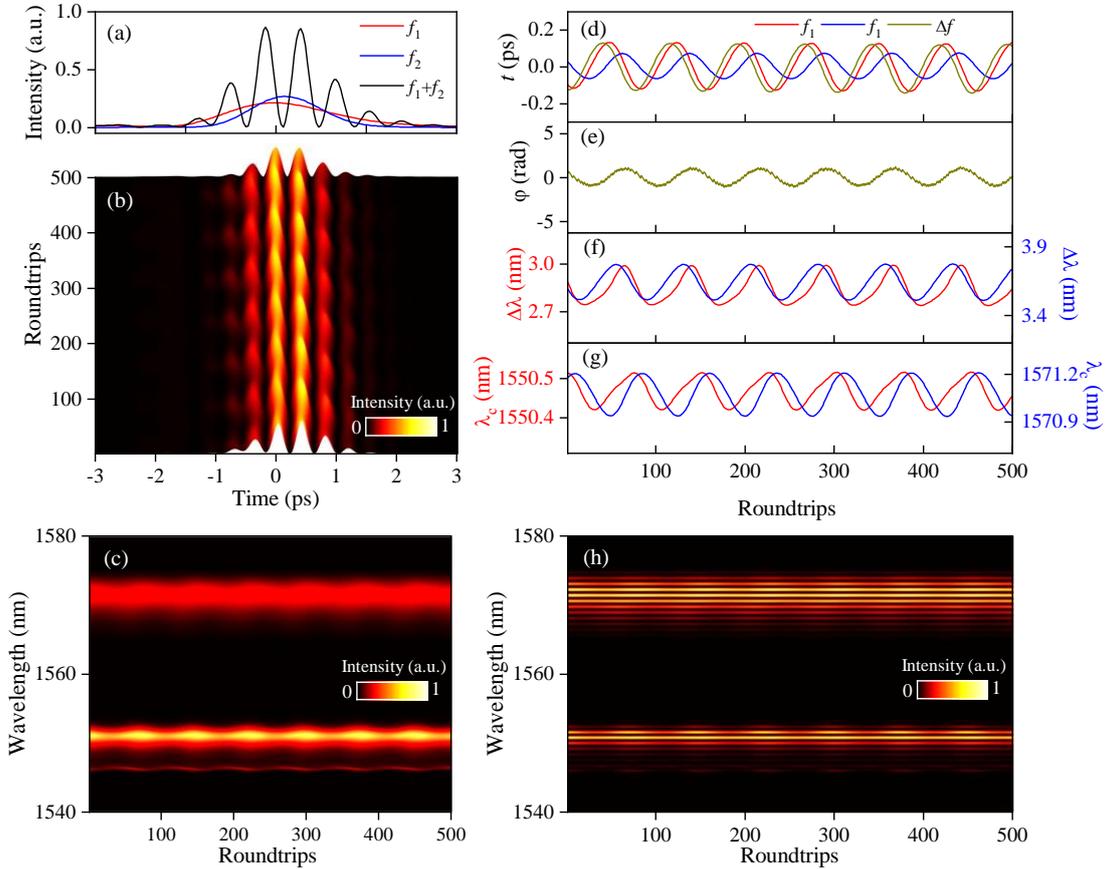

Fig. 4. The DBM dynamics produced by simulations. (a) Pulse profiles of DBM ($f_1$+$f_2$) and individual breathers ($f_1$ and $f_2$) are obtained at the propagation distance corresponding to 500 RTs in (b). The temporal (b) and spectral (c) DBM evolution vs. the number of RTs. (d) The evolution of the temporal separation between the pulses, $\Delta f$, and positions of the breathers, $f_{1,2}$. (e) The evolution of the phase difference between the breather components. (f) The evolution of bandwidths and (g) central wavelengths of the bound breathers. (h) The evolution of interference spectra, as produced by RDIS applied to the simulation results displayed in (c).

The variation of temporal and spectral features of DBM can be extracted from Figs. 4(b) and 4(c). The periodic change of the separation between the bound pulses, displayed in Fig. 4(d), represents intrinsic vibrations of DBM, as predicted by the simulations. Note that the separation, $\sim 100$ fs, is much smaller than the pulse width, $\geq 500$fs. The evolution of positions of the breathers is similar to the experimental observations in Fig. 2(h), in the form of the out-of-phase trajectories. The evolution of the phase difference between the breathers, as well as their spectral widths and central wavelengths, produced by the simulations in Figs. 4(e), 4(f)and 4(g), is similar to the experimentally observed oscillations of the same variables in Figs. 2(i), 2(e) and 2(d). The oscillations of the breathers, as predicted by the simulations, are mutually out-of-phase, in agreement with the experiment. The experimentally observed evolution is somewhat more complex than produced by the simulations, due to noise effects, as shown in Supplemental Material [65]. The predicted oscillation period and amplitude are different from the experimental results because the model cannot take into account all experimental peculiarities, such as an exact form of the fiber gain and higher-order GVD in the fiber segments.

DBM is represented here by the periodic solution under the condition of the quasi-group-velocity matching, which is distinct from previous works [51–57,60]. It is supported by an effective XPM-induced binding potential, cf. Ref. [52]. This is different from the single-color setting where solitons interact via the SPM-induced force, which depends on the phase shift between the solitons [39-42]. Intrinsic oscillations of the single-color soliton molecule are represented by periodic solutions of the nonlinear system. The out-of-phase oscillations can be produced when soliton features at two wavelengths are different. The out-of-phase oscillations of the central wavelength and temporal position, which are typical dynamical regimes for DBMs in mode-locked fiber lasers, can be explained by a model based on coupled harmonic oscillators [70]. This model is demonstrated to produce out-of-phase and in-phase oscillations with suitable parameters in Supplemental Material [65].

To expand the findings, RDIS was also applied to the relative motion and phase evolution in a static DBM, as shown in Fig. S2 in Supplementary Material [65]. No periodic variations were observed in that case, and the spectral pattern stayed constant. RDIS was also applied to the simulation results presented in Fig. 4(c), see Fig. S5 in Supplementary Material [65]. The so generated evolution pattern, which is plotted in Fig. 4(h), is identical to the experimental results in Fig. 2(f), and the extracted evolution of the pulse separation and phase agree well with Figs. 4(d) and 4(e). No obvious variation of the relative phase is observed in the interference pattern of Figs. 2(f) and 4(h), as the position of the interference peak does not change significantly. This finding can be explained by the fact that the spectral profile of the SM is determined not only by the pulse separation and phase shift, but is also affected by the wavelength difference. In Fig. S6 of Supplementary Material [65], RDIS is applied to the pulses with the periodically varying central wavelength, intensity and phase. The calculations reproduce the spectral evolution similar to Fig. 4(h). The results corroborate that the experimental and theoretical evolution of the phase shift, shown in Figs. 2(h) and 4(e), respectively, originate from the variation of the central wavelength. Therefore, the evolution of the central wavelength in Fig. 2(d) corresponds to the relative phase in Fig. 2(j). The dependence of dynamical parameters of DBMs on the pump power is also verified theoretically in Supplemental Material [65].

In conclusion, bound states of solitons in the form of DBMs (dichromatic breather molecules) have been experimentally created and explored in the synchronized mode-locked fiber laser by means of RDIS (real-time delay-shifting interference spectrum) and real-time spectroscopy techniques. The breathing behavior was observed in the real-time spectral evolution. The variation of the temporal separation and phase shift between the pulses were extracted by means of RDIS. Thus, the intrinsic vibrations of the DBM were revealed by the evolution of these characteristics. The breathing ratio and vibration amplitude vary along with the pump power. The evolution of the DBM features out-of-phase trajectories of the constituent breathers. Numerical simulations explain the creation of the DBMs in the framework of the model combining oscillations of the population inversion and the system of the XPM-coupled CGLEs. Simulations agree well with the experimental observations, verifying the relation between the Q-switching instability and appearance of breathers in mode-locked lasers. The methods elaborated in this work can be applied to other settings. In particular, it is natural to consider bigger dichromatic complexes, built of two breathers carried by one wavelength and one or two breathers carried by the other [71].

## Acknowledgements

The authors acknowledge support from the National Natural Science Foundation of China under Grant Agreements 61705193 and 62205296, the Natural Science Foundation of Zhejiang Province under Grant Nos. LGG20F050002, LQ23F050004 and LY19F050014, the Open Project Program of Wuhan National Laboratory for Optoelectronics No. 2020WNLOKF008, the Independently Project by Zhejiang Normal University under Grant No. 2021ZS05, and Israel Science Foundation through grant No. 1695/22.